\documentclass[fleqn,twoside]{article}
\usepackage{espcrc2}

\usepackage{graphicx}

\def\Agata{R\' o\. za\' nska~}

\title{Universal spectral shape of AGN with high accretion rate}

\author{Marek Niko\l ajuk\address[CAMK]{Nicolaus Copernicus Astronomical 
        Center, Bartycka 18, 00-716 Warsaw, Poland}%
        \thanks{Send offprint requests to M. Niko\l ajuk, mark@camk.edu.pl},
        Bo\.zena Czerny\addressmark,
        Agata R\'o\.za\'nska\addressmark,
        and 
        Anne-Marie Dumont\address[DAEC]{Observatoire de Paris-Meudon,
        DAEC, Meudon, France}}

\begin{document}

\begin{abstract}
The spectra of radio quiet and NLS1 galaxies show suprising similarity
in their shape. They seem to scale only with the accretion rate but
not with central black hole mass. We consider two mechanisms modifying
the disk spectrum.  First, the outer parts of the disk are irradiated
by the flux emerging from the inner parts. This is due to the
scattering of the flux by the extended hot medium (warm absorber).
Second process is connected with the development of the disk warm
Comptonizing skin above the disk and/or coronae.  Our scenario applies
only to object with relatively high luminosity to the Eddington
luminosity ratio for which disk evaporation is inefficient.
\end{abstract}

\maketitle

\section{Introduction}

\begin{figure}[t]
\includegraphics[width= 0.42\textwidth, bb=50 190 530 650]{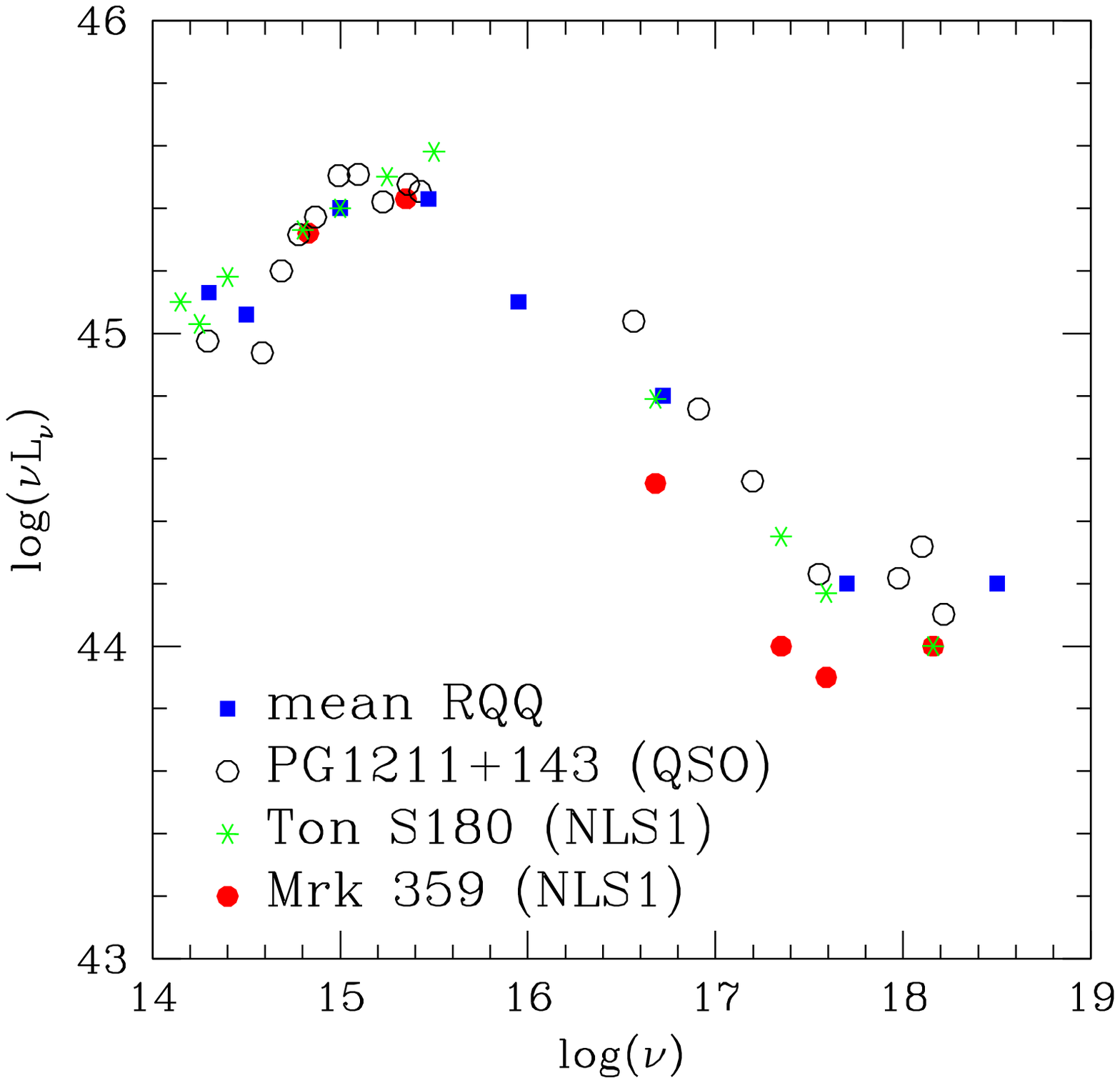}
\caption{
The broad band spectra of quasars: composite spectrum of RQQ (filled
squares), PG 1211+143 (open circles) and NLS1 galaxies: Ton S180
(stars), Mrk 359 (filled circles).  The spectra of the last three
objects are shifted by 0.1, 0.6 and 2.0, respectively, to coincide at
the B band.  
}
\label{exmpl}
\end{figure}

Radio quiet quasars and Narrow Line Seyfert 1 galaxies are the sources
expected to accrete mass at a high rate (e.g. ref. \cite{coll02}).
The spectra of those objects show similarity in their shape
(Fig.~\ref{exmpl}).  The idea of the universal spectral shape of those
galaxies appears in the literature since many years
(e.g. ref. \cite{wafin93}). It is well known that the simplest
standard Shakura-Sunyaev disk model, under assumption of the local
black body emission, cannot account for the observed X-ray spectra of
AGN. The model is also not quite satisfactory in the optical/UV band.

We propose two modifications to the standard model which directly
result from our knowledge of the accretion disk physics and an AGN
surrounding.

\section{Irradiation by warm absorber}

\begin{figure}
\includegraphics[width= 0.45\textwidth, bb=30 292 570 695]{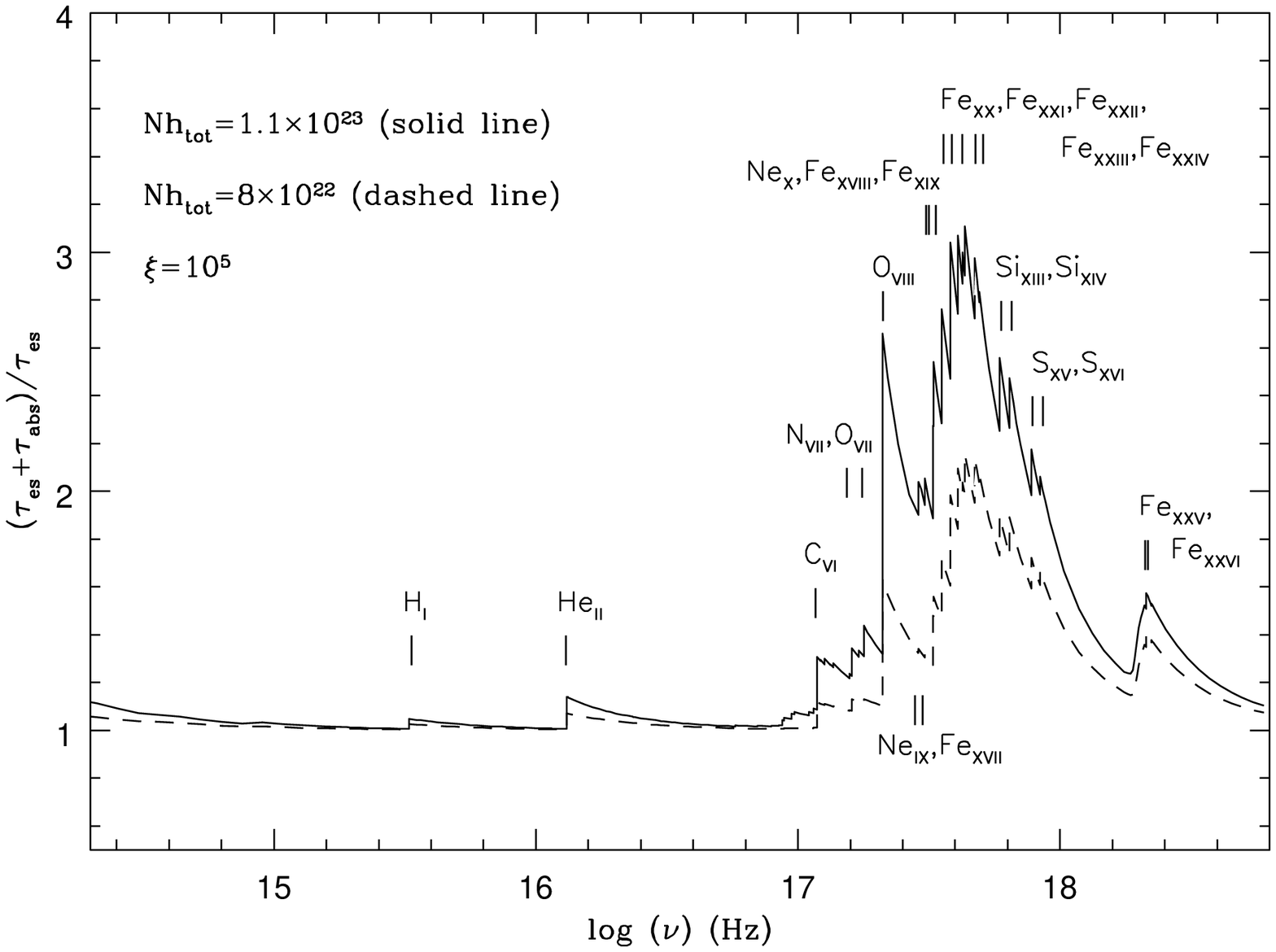}
\caption{
The ratio of the total optical depth (absorption + scattering) to the 
Thomson optical depth (scattering) as a function of energy for the
warm absorber model in Ton S180: Total column density $1.1 \times
10^{23}$ cm$^{-2}$ (solid line) and $8 \times 10^{22}$ cm$^{-2}$
(dashed line), ionization parameter at the surface $\xi = 10^{5}$.
Density profile of the warm absorber comes from the constant pressure
condition.
}
\label{kabs}
\end{figure}

The warm absorber is a significantly ionized medium present in our
line of sight towards sources. It produces narrow absorption lines and
absorption continua in the soft X-ray band. The medium temperature is
of order of $10^5$-$10^6$ K. The column density is frequently of order
of $10^{23}$ cm$^{-2}$ or more. The warm absorber is extended but
probably clumpy (e.g. ref. \cite{krol02}). Most of the radiation from
the nucleus of the considered objects is emitted in the frequency
range $10^{15}$-$10^{17}$ Hz.  In this range the electron scattering
dominates the total crossection. The warm absorber actually
predominatly scatters the incoming photons (Fig.~\ref{kabs}) instead
of absorbing them (ref. \cite{roz03}).

\begin{figure}
\begin{center}
\includegraphics[width= 0.27\textwidth, bb=50 20 710 795]{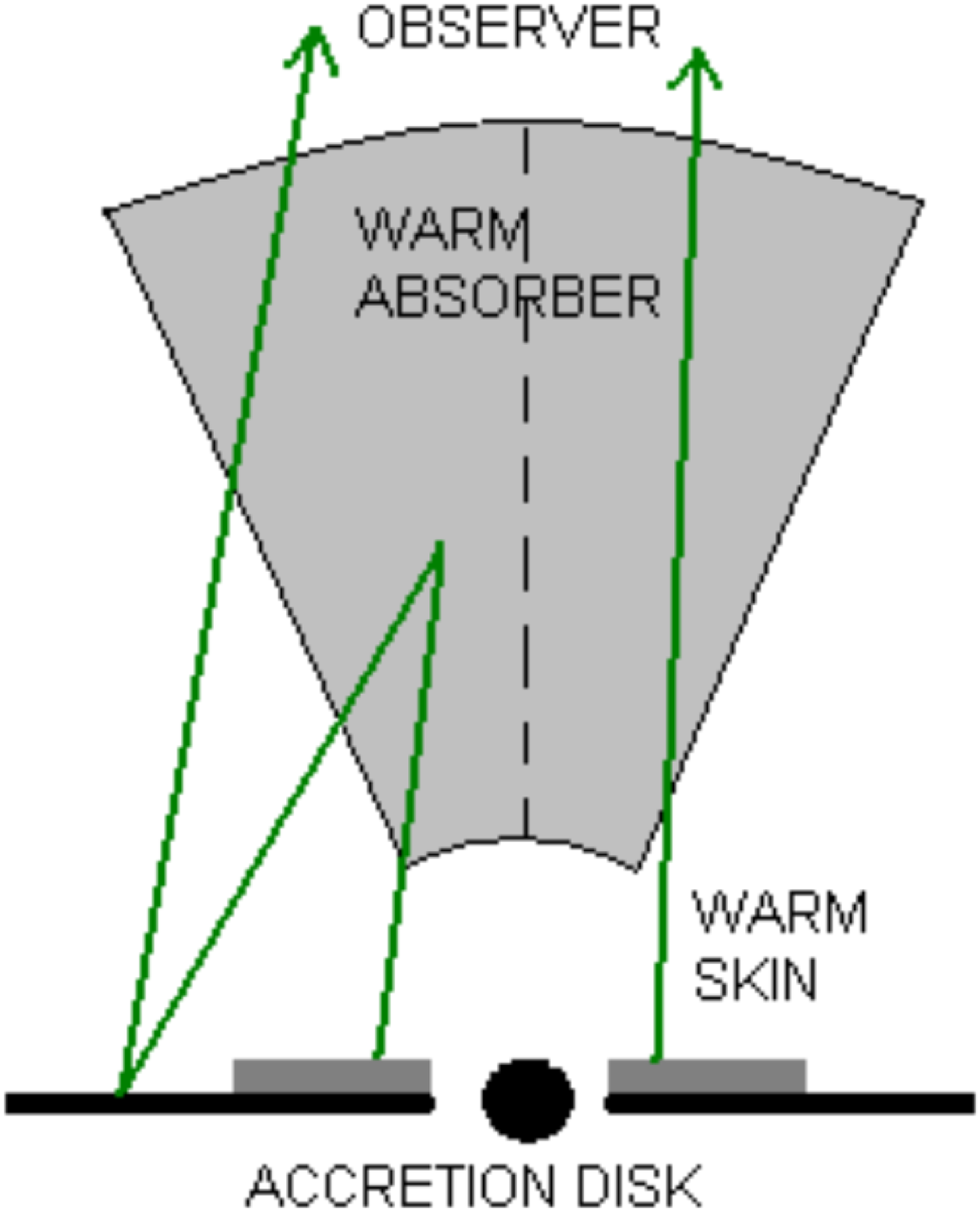}
\end{center}
\caption{
The schematic picture of a bright AGN with a warm absorber 
and a warm disk skin.
}
\label{wasp}
\end{figure}

We envision the geometry in Fig.~\ref{wasp}. We neglect the clumpiness
of the medium and assume that it is located predominantly along the
symmetry axis, $\cos i \approx 1$.  We parameterize the region by the
total optical depth for electron scattering, $\tau_{wa}$, minimum
$z_{\rm min}$ and maximum $z_{\rm max}$ distance of the warm absorber
from the nucleus.  We assume that the density profile of the warm
absorber is $\rho(z) \propto z^{-\delta}$.  For purpose of the rough
estimate of modifying disk spectrum we do not introduce any specific
dependence on the distance from the symmetry axis. The irradiation of
the disk at a distance $r$ from the center can be approximately
represented as
\begin{equation}
F_{irrad} = {2L \tau_{wa} f_{wa}(\delta - 1) \over z_{min}^{1 - \delta} 
- z_{max}^{1 - \delta}} {\int_{z_{min}}^{z_{max}} {z^{-\delta + 1} \over 
(z^2 + r^2)^{3/2}} dz}
\label{eq:irrad}
\end{equation}
where $L$ is the bolometric luminosity of an accretion disk and 
$f_{wa}$ is the fraction of emission from the nucleus passing through the 
warm absorber. 

The warm absorber modifies the emerged spectrum (predo\-minatly at
longer wavelengths -- optical and the IR band). The effect is only weakly 
influenced by $z_{\rm max}$ and the density distribution $\rho(z)$.
More important is the exact location of the inner edge, $z_{\rm min}$.
The exemplary spectrum is are shown by dashed line in Fig.~\ref{decompo}. 

\section{Warm Comptonizing skin}

Comptonization of a fraction of disk photons is clearly needed in
order to reproduce the soft X-ray part of the spectrum.  In the high
luminosity objects the disk extends down to the marginally stable
orbit and some warm skin covers the inner region. Similar
interpretation holds both for galactic black holes in their High State
or Very High State and for luminous AGN (ref. \cite{bcz02,gier01}).
Here we propose a natural constraint for the properties of this warm
skin based on stability analysis in the radiation pressure dominated
region of the disk.

We assume that the accretion disk is azimuthaly divided into two
layers: inner cold part and a warm Comptonizing skin (Fig.~\ref{wasp}).
We futher assume that fraction of the energy is transported from the
disk interior towards the skin as a magnetic flux. In this way proceed
the heating of the warm skin. 
Therefore, the energy balance of the disk interior reads:
\begin{equation}
Q^{+} = Q^{-} + W \, ,
\label{eq:balance}
\end{equation}
where 
$Q^{+} = {3 \over 2} \alpha P H \Omega_K $ 
is the energy flux dissipated in the disk interior,
$Q^{-} = {4 \sigma_{B} T^4 \over 3 \Sigma \kappa_{es}}$ 
is the radiative cooling of the disk interior. 
$W$ is the magnetically transported flux and 
$W \propto B^2 v_A H \propto P^{3/2} H^{3/2}$ 
($B^2$ -- the energy density of the magnetic field,
$v_A$ -- the Alfven velocity, $H$ --
the disk thickness, $P$ -- the pressure). The warm
skin is heated by flux tubes which cover only a fraction of the
boundary surface.  We assume that this covering is proportional to the
disk thicknes, $H$.

The disk is thermally stable if:
\begin{equation}
  \delta Q^+ < \delta Q^{-} + \delta W \, .
\label{eq:stab1}
\end{equation} 
Our calculations show that the presence of magnetic transport
stabilizes the disk since its dependence on the temperature is steep
enough (the same as in the case of advection).  When $w \equiv
W/Q^{+} = 0.5$ the disk already becomes stable even if strongly
dominated by the radiation pressure.  We therefore consider it
possible that the disk in a natural way develops this magnetically
mediated transport in the radiation pressure dominated region.
This region of the disk would be therefore covered with a warm skin
and the transport efficiency saturates at the value $w = 0.5$\,.

The skin cools predominantly by Comptonization. The cooling
efficiency is fixed by the parameter $w$, i.e. the Compton
amplification factor, $A$, is given by
\begin{equation}
A= {Q^{-} + W \over Q^{-}} ={1 \over 1 - w} = 2 \quad (\mathrm{for}\ w=0.5).
\end{equation}

Therefore, the condition $A = 2$ roughly determines the slope,
$\alpha_{E}$, of the soft X-ray spectrum. It provides  a constraint for
the combination of the skin optical depth, $\tau_{skin}$, and its
temperature, $T_{skin}$.

\begin{figure}[tb]
\includegraphics[width= 0.42\textwidth, bb=50 190 530 650]{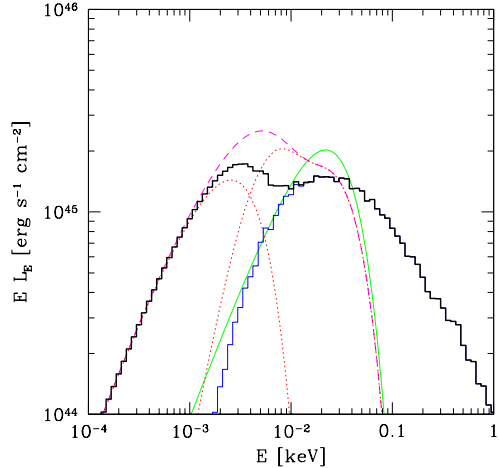}
\caption{
The spectrum for the mean RQQ. The standard disk model 
(thin continuous line). The disk is affected by the warm absorber 
through irradiation (dashed lines). Disk is divided into inner and 
outer part (dotted lines). The warm skin modify the emerged inner disk 
radiation (continuos thin histogram). The total spectrum 
(continuous thick histogram).
}
\label{decompo}
\end{figure}

\begin{figure}
\includegraphics[width= 0.42\textwidth, bb=50 190 530 650]{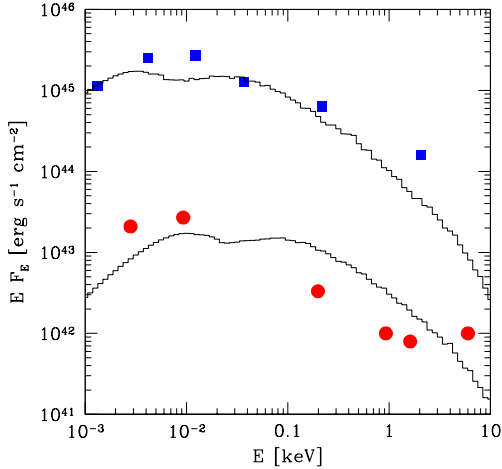}
\caption{
The exemplary accretion disk models (continuous histograms), the
spectrum of composite spectrum of quasars (filled squares) and Mrk
359 (filled circles).  
}
\label{sp}
\end{figure}

The role of the irradiation and the Comptonization by the warm
absorber and disk skin is illustrated in Fig.~\ref{decompo}. The model
does not fit exactly the observed data points (Fig.~\ref{sp}) since our
aim of our model is to reproduce in a possibly natural way the scaling
of the spectra with mass and not fitting.

\begin{center}
\begin{table}[htb]
\normalsize
\caption{Summary of model parameters.}
\label{tab.1}
\begin{tabular}{lrr}
\hline
Parameter & quasar composite &  Mrk 359 \\
\hline
$M [M_{\odot}]$       & $1.0 \times 10^8$ & \quad $1.0 \times 10^6$ \\
$\dot m [L/L_{Edd}]$  & 0.3  & 0.3  \\
$\tau_{wa}$           & 0.25 & 0.25 \\
$f_{wa}$              & 0.5  & 0.5  \\
$z_{min} [R_{Schw}]$  & 100  & 100  \\
$z_{max} [R_{Schw}]$  & 10 000 & 10 000\\ 
$A$                   & 2    & 2    \\
$\tau_{skin}$         & 3.32 & 3.32 \\
$T_{skin}$ [keV]      & 3    & 3    \\
$r_{skin} [R_{Schw}]$ & 300  & 300  \\
\hline 
\end{tabular}
\end{table}
\end{center}

\section{Conclusion}

We develop a model which is appropriate for high luminosity objects
like quasars and NLS1 galaxies. 

We propose two mechanisms which together may be responible for the
relatively uniform spectral shape of the AGN luminosities in term of
$L/L_{Edd}$. First mechanism is back-scattering of the
light emerged from disk by warm absorber.  Second mechanism is the
Comptonization in the marginally stable accretion disk warm skin.

Main conclusions of our investigation are following:
\begin{itemize}
\item Irradiation results in softer/UV spectra than predicted by classical 
disc models.
\item The warm skin stabilizes the radiation pressure dominated disk.
\item It Comptonizes the disk emission and the soft X-ray part of the 
spectrum is produced.
\item The condition of stability leads to a specific value of the
Compton amplification factor ($A=2$) which roughly determines the soft 
X-ray slope. 
\item Both effects lead to the radiation spectrum which shape does not
strongly depend on the parameters.
\end{itemize}

Our results are in agreement with the conclusion of
ref.~\cite{woourry02} that there is no strong correlation between the
Eddington ratio and the mass.

Lower luminosity AGN (in terms of $L/L_{Edd}$), like Seyfert
galaxies, are not expected to be represented by the discussed
model. In those sources the disk evaporation is most probably very
effective, as expected from the models
(e.g. ref. \cite{rozbcz00,liu02}), and the inner disk is replaced with
a hot, possibly advection dominated flow (e.g. ref.~\cite{nayi94}).

\vspace{1em}
\emph{Acknowledgements}
Part of this work was supported by grants 2P03D00322 and
PBZ-KBN-054/P03/2001 of the Polish State Committee for Scientific
Research, and Jumelage/CNRS No.~16 ``Astronomie France/Pologne''.


\end{document}